\documentclass[aps,prl,twocolumn,showpacs,amsmath,amssymb]{revtex4-1}
\usepackage{graphicx}% Include figure files
\usepackage{dcolumn}% Align table columns on decimal point
\usepackage{epstopdf}
\usepackage{bm}% bold math
\usepackage{bbold}

\begin{document}
\title{Orbital-selective conductance of Co adatom on the Pt(111) surface}
\author{V.V. Mazurenko$^{1}$, S.N. Iskakov$^{1}$, M.V. Valentyuk$^{1}$, A.N. Rudenko$^{1,2}$ and A.I. Lichtenstein$^{3}$}
\affiliation{$^{1}$Theoretical Physics and Applied Mathematics Department, Ural Federal University, Mira Str.19,  620002
Ekaterinburg, Russia \\
$^{2}$ Institute of Chemical Reaction Engineering, Hamburg University of Technology, Eissendorfer Str. 38, 21073 Hamburg, Germany \\
$^{3}$Institute of Theoretical Physics, University of Hamburg, Jungiusstrasse 9, 20355 Hamburg, Germany}
\date{\today}

\begin{abstract}
We propose an orbital-selective model for the transport and magnetic properties of the individual Co impurity deposited on the Pt(111). Using the combination of the Anderson-type Hamiltonian and the Kubo's linear response theory we show that the magnetization and $dI/dV$ spectrum of Co adatom are originated from the $3d$ states of the different symmetry. A textbook expression for the spin-dependent differential conductance provides a natural connection between magnetic and transport properties of Co/Pt(111). We found that it is possible to detect and to manipulate  the different $3d$ states of the Co adatom by tuning the tip-impurity distance in STM experiments.
\end{abstract}

\maketitle

Spin-polarized scanning tunneling microscopy (SP-STM) and spectroscopy (SP-STS) are important tools of the modern physics for studying electronic and magnetic properties of a transition metal adatom deposited on a surface \cite{WiesenRMP}.
Depending on the surface properties that can be insulating \cite{otte} or metallic \cite{Meier}, magnetic \cite{Lounis} or nonmagnetic \cite{Jamneala},  the STM conductance spectra demonstrate different types of the spin-dependent excitations such as Kondo resonance \cite{Madhavan}, step-type excitations, quasiparticle peak \cite{Mazurenko}. The resulting $dI/dV$ conductance spectra can be well reproduced in the framework of the localized (Heisenberg-type)  \cite{Appelbaum, Anderson, Fransson, fernandez, rudenko} or itinerant (Stoner-type) spin-model approaches \cite{Wiesendanger}.

One of the fascinating results of the STM experiments is the possibility to manipulate and to control the individual spin as well as the simplest nanostructures.
For example, the authors of Ref.\cite{hirjibehedin} were able to construct magnetic chains of different lengths which displayed evidence of coupled-spin behavior.
In turn Seratte $et$ $al.$ \cite{Seratte} have demonstrated a direct access to the spin direction of individual atoms.  

Recently, the $3d$-orbital degree of freedom has attracted a lot of attention due to its important role in the massive variations of the magnetic and transport properties of strongly correlated systems \cite{Chakhalian, Khaliullin}. For instance, using  X-ray reflectivity data the authors of Ref.\cite{Benckiser} have proposed a method to derive the orbital polarization of transition-metal oxide interfaces.  In case of a transition metal adatom deposited on a surface the STM technique can be applied to probe and to manipulate  the individual $3d$ states. For this purpose one should use a system that demonstrates a strong variation of the electronic and magnetic properties depending on the symmetry of the specific $3d$ orbital. Such a variation results in a strong orbital polarization of the STM spectra.

In this paper we propose the Co impurity deposited on the Pt(111) surface to be one of the best candidate to probe the individual $3d$ states using STM technique since it demonstrates a strong orbital polarization of the $dI/dV$ spectrum. Using the exact diagonalization of the many-body electronic Hamiltonian and the Kubo's linear response theory we show that the main contributions to the magnetization curve and the peak of the spin-polarized $dI/dV$ spectrum are due to the $3d$ states of the different symmetry. Based on the obtained results we conclude that one can detect the different $3d$ states by varying the tip-impurity distance. 
 
\begin{figure}[]
\includegraphics[width=0.17\textwidth,angle=0]{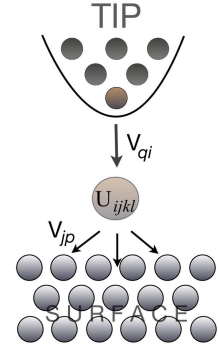}
\caption{(Color online) Schematic  representation of the computer simulation setup used in this work. The electrons hop from the tip to the surface via the impurity with the full Coulomb interaction matrix.}
\label{structures}
\end{figure}  
    
{\it Computer simulation setup.} To study the transport properties of the Co/Pt(111) system we construct the Anderson-type Hamiltonian that contains tip, impurity and surface states (Fig.1)

\begin{eqnarray}
H = \sum_{q \sigma} \epsilon_q^{\sigma} b^{+}_{q \sigma} b_{q \sigma} + \sum_{p \sigma} \epsilon_{p} c^{+}_{p \sigma} c_{p \sigma} 
+ \sum_{i \sigma}   (\epsilon_{i} - \mu) n_{i \sigma}  \nonumber \\
+ \frac{1}{2} \sum_{i}g \mu_B B^z (n_{i \uparrow} - n_{i \downarrow}) + \frac{1}{2} \sum_{\substack {ijkl\\ \sigma \sigma'}}  U_{ijkl} d^{+}_{i \sigma} d^{+}_{j \sigma'} d_{l \sigma'} d_{k \sigma} \nonumber \\
+  \sum_{i p \sigma} ( V_{i p} d^{+}_{i \sigma} c_{p \sigma} + H.c.) 
+ \sum_{i q \sigma} ( V_{iq} d^{+}_{i \sigma} b_{q \sigma} + H.c.).  
\end{eqnarray}
Here $\epsilon_{i}$, $\epsilon_{q}^{\sigma}$ and $\epsilon_{p}$ are energies of the impurity, tip and surface states, $d^{+}_{i \sigma}$ and $b^{+}_{q \sigma}$ ($c^{+}_{p \sigma}$) are the creation operators for impurity and tip (surface) electrons, $V_{iq}$ ($V_{ip}$) is a hopping between impurity and tip (surface) states, $U_{ijkl}$ is the Coulomb matrix element and the impurity orbital index $i$ ($j$, $k$, $l$) runs over the $3d$ states ($xy, yz, 3z^2 - r^2, xz,x^2-y^2$). In what follows we assume $V_{iq}$ and $V_{ip}$ are real. Using finite-temperature exact diagonalization method \cite{Capone} we calculated one-particle and two-particle correlation functions for 40 excited states, which is enough to work at the experimental temperature of  0.3 K with the Boltzmann factor smaller than 10$^{-6}$. 

The energies of the impurity and surface orbitals as well as the impurity-surface hybridization parameters can be estimated from the first-principles calculations by fitting procedure described in Ref.\onlinecite{Mazurenko}. The on-site Coulomb and intra-atomic exchange interactions were estimated to be $U$ = 6.75 eV and $J$ =0.9 eV \cite{Mazurenko}. Since the spin-polarization of the tip effective orbital and its overlapping with the impurity wave functions are unknown, they are free parameters in our model that should be carefully chosen by using scientifically sounded arguments. 

The tip-impurity hopping process can be simulated by three z-oriented adatom orbitals that overlap with the tip orbitals. To take into account all the excitation paths between the tip and adatom we define one complex Wannier function centered on the lowermost tip atom. This Wannier orbital overlaps with $yz$, $xz$ and $3z^2-r^2$ orbitals of the cobalt impurity.  We used a Slater-parametrization \cite{Slater} to define the corresponding hopping integrals, $V_{q \, 3z^2-r^2} = 2 V_{q \, yz} = 2 V_{q \, xz}$.
Thus the number of the adjustable parameters can be reduced to two items, the spin polarization of the tip and the tip-impurity hybridization. It is useful to define some bounds for these parameters.

Since in the STS experiments \cite{Meier} the tip with out-of-plane polarization was used then the magnetization is an intrinsic property of the tip and it should be considerably larger than the external magnetic field ranging from  -7 T to 7 T.  The tip-impurity hybridization parameters strongly depend on the vertical position of the tip. It is natural to assume that the hybridization between tip and adatom should be considerably smaller than that between adatom and surface orbitals (0.3 - 0.7 eV).  Thus we have solved the model Hamiltonian with the tip-impurity hopping integral ($V_{q \, 3z^2-r^2}$) ranging from 50 meV (weak hybridization) to 150 meV (strong hybridization). Within the Slater-parametrization \cite{Harrison} it corresponds to the tip-impurity distances of 4.1 \AA \, and 3.2 \AA. The spin polarization of the tip $\Delta = \epsilon_{q \uparrow} - \epsilon_{q \downarrow}$ ranges from 10 meV (weak polarization) to 100 meV (strong polarization). In what follows we assume that the spin polarization of the tip is along z-axis.  

{\it Magnetization of impurity states.} According to the STM experiments \cite{Meier}, the differential conductance spectrum of the individual Co impurity as a function of the external magnetic field demonstrates an S-shaped curve form.  Since there are no signs of hysteresis, the applied magnetic fields are not sufficient  to induce magnetism. To reproduce this situation we have performed model calculations with the magnetic field $\vec B$ ranging from -25 T to 25 T and different values of $V_{q \, 3z^2-r^2}$ and $\Delta$. In all the cases we observed the similar behavior of the magnetization curves of the different orbitals.  
The typical example is presented in Fig.2.  One can see that all the $3d$ states can be classified with respect to the sensitivity to the magnetic filed. The orbitals of $yz$ and $xz$ symmetry are almost paramagnetic and show a weak response to the field. At the same time the slope of the $3z^2-r^2$ curve is the largest one, which is a result of a strong magnetic moment localization \cite{Mazurenko}. 

\begin{figure}[]
\includegraphics[width=0.43\textwidth,angle=0]{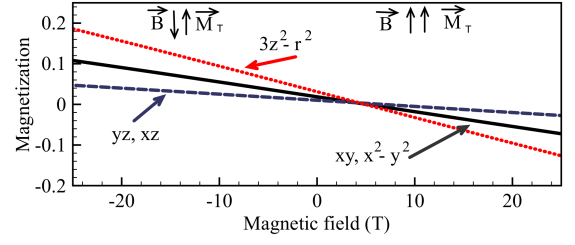}
\caption{ (Color online) Magnetization (in $\mu_B$) of the impurity states as a function of the external magnetic field, $B^z$ calculated with $\Delta$ = 10 meV and $V_{q \, 3z^2-r^2}$ = 100 meV. $\vec M_{T}$ is the magnetization of the tip.}
\label{structures}
\end{figure} 

{\it Spin-dependent conductance.} Since the Hamiltonian Eq.(1) is diagonalized the obtained eigenvalues and eigenfunctions can be used in order to calculate
complex two-particle correlation functions. The most important one is the differential conductance that in the framework of the Kubo's linear response theory can be written 
\begin{eqnarray}
\frac{dI}{dV} (\omega, \Delta, \vec{B}) = \frac{1}{Z \omega} \sum_{n n'} \frac{j_{nn'} j_{n'n}}{\omega + E_{n} - E_{n'}} [e^{- \beta E_{n}} - e^{- \beta E_{n'}}],
\end{eqnarray}
where  $Z$ is the partition function, $E_{n}$ is the eigenvalue of the Hamiltonian Eq.(1) and $j_{nn'}$ is a matrix element of the current operator. To avoid the extrapolation from Matsubara frequencies to real frequencies we use energy with a small imaginary part $\omega + i \delta$.   

In our finite-cluster approach the current operator can be derived from the quantum Hamiltonian, Eq.(1) by using an approach proposed in Ref.\cite{Kubo} 
\begin{eqnarray}
j = \frac{i}{2}  \sum_{i q p \sigma} [V_{iq} (d^{+}_{i \sigma} b_{q \sigma} -  b^+_{q \sigma} d_{i \sigma})
+  V_{i p} (d^{+}_{i \sigma} c_{p \sigma} -  c^+_{p \sigma} d_{i \sigma})].
\end{eqnarray}
The current operator has a symmetrized form and contains the average of the tip-impurity and impurity-surface contributions. 
Such an operator can be used to calculate the expectation value of the current passing through the impurity in different regimes, i.e. in the linear response or in the steady state.  A similar symmetrization of the current contributions was used in the previous investigations, for instance in Ref.\cite{Wingreen}.
  
The resulting differential conductance, Eq.(2) contains the sum over the spin and effective orbital indices.
Let us first discuss the orbital dependence of the $dI/dV$ spectrum.
All the excitations of the same spin, $ j_{nn'}^{\sigma} j_{n'n}^{\sigma}$ can be classified into the direct and indirect processes.
The examples of the direct excitations are presented in Fig.3(a). There are a tip-impurity-surface excitation (process I)
$\langle n |c^{+}_{p \sigma} d_{i \sigma} d^{+}_{i \sigma} b_{q \sigma} | n \rangle$,
a tip-impurity-tip excitation (process II) $\langle n |b^{+}_{q \sigma} d_{i \sigma} d^{+}_{i \sigma} b_{q \sigma} | n \rangle$
and a surface-impurity-surface excitation (process III) $\langle n |c^{+}_{p \sigma} d_{i \sigma} d^{+}_{i \sigma} c_{p \sigma} | n \rangle$. 
Here $q$ and $p$ denote tip and surface effective orbitals, respectively. These direct excitations can be reduced to the form that contains the spin operator of the corresponding impurity orbital, therefore, one should expect that these 
contributions are very sensitive to the applied magnetic field. 
\begin{figure}[b]
\includegraphics[width=0.45\textwidth,angle=0]{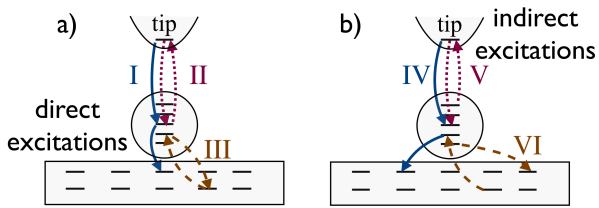}
\caption{ (Color online) Schematic representation of the direct (process I, II, III) and indirect (process IV, V and VI) excitations that contribute to the differential conductance.}
\label{structures}
\end{figure}

It is not the case for the indirect excitations [Fig.3(b)] described by 4-point correlation functions
which are
$\langle n |c^{+}_{p \sigma} d_{j \sigma}  d^{+}_{i \sigma} b_{q \sigma} | n \rangle$
for the tip-impurity-impurity-surface transition (process IV), 
$\langle n |b^{+}_{q \sigma} d_{j \sigma}  d^{+}_{i \sigma} b_{q \sigma} | n \rangle$
for the tip-impurity-impurity-tip transition (process V), 
and 
$\langle n |c^{+}_{p' \sigma} d_{j \sigma} d^{+}_{i \sigma} c_{p \sigma} | n \rangle$
for the surface-impurity-impurity-surface excitation (process VI). 
Such processes are due to the full Coulomb interaction matrix taken into account for the impurity states.
We believe that they play an important role for simulating surface-assisted Hall effects \cite{Hall} and left their analysis for a future investigation. 

Having analyzed the different types of excitations we are now in a position to connect the finite-cluster conductance with that experimentally observed \cite{Meier}. 
For this purpose we should choose the terms that correspond to the tip-impurity-surface excitations (the process I for $3z^2-r^2$, $yz$ and $xz$ orbitals).  However, these excitation processes are described by the two-particle Green's function that is non-diagonal with respect to the site indices. It means that the spectral function can change the sign and, as a consequence, the resulting conductance can be negative. To overcome this problem we symmetrized the Green's function of the process I \cite{sym}. Thus the spin-diagonal part of the STM conductance takes the form   
\begin{eqnarray}
\frac{dI_{STM}}{dV} (\omega, \Delta, \vec{B}) =  \frac{1}{4\omega Z} \sum_{i p \sigma} \sum_{n n'} V_{iq} V_{ip} (e^{- \beta E_{n}} - e^{- \beta E_{n'}})\nonumber \\
\times [\frac{ |\langle n' | d^+_{i \sigma} b_{q \sigma} + d^+_{i \sigma} c_{p \sigma}  |n \rangle|^2 } {\omega + E_{n} - E_{n'}} + 
\frac{ |\langle n' | b^+_{q \sigma} d_{i \sigma} + c^+_{p \sigma} d_{i \sigma}  |n \rangle|^2 } {\omega + E_{n} - E_{n'}}].
\end{eqnarray}

As for the non-diagonal spin contributions to the differential conductance, $ j_{nn'}^{\sigma} j_{n'n}^{\sigma'}$ ($\sigma \neq \sigma'$) that can be calculated by using Eq.(2) they correspond to the spin-flip-assisted tunneling processes \cite{fernandez}. Their contribution to the STM signal will be discussed below. 

The spin dependence of the conductance provides an access to the magnetic properties of the surface nanosystem. 
To describe the connection between conductance and magnetic moment of the impurity a classical Heisenberg-type expression was proposed by Wortmann $et$ $al.$ \cite{Blugel}. Within this approach the spin-dependent part of the conductance is proportional to the product $dI_{STM}/dV \sim \vec M_{T} \vec M_{Co}$, where $\vec M_{T}$ and $\vec M_{Co}$ are the magnetization of the tip and Co adatom, respectively. 

In turn, Fransson $et$ $al.$ \cite{Fransson} and Delgado $et$ $al.$ \cite{Fernandez2} have shown that the tunneling conductance contains a background contribution that is independent of the local spin, a dynamical conductance which is proportional to the local impurity spin, and a susceptibility contribution that corresponds to the spin-spin correlation function of the local spins. Using such an approach one can extract the local spin moment and analyze the spin excitations recorded in conductance measurements. 

We would like to stress that the previous considerations of the spin-dependent conductance are based on some approximations, which strongly limits their applicability. The crucial one is the absence of the quantum fluctuations between the tip, impurity and surface states. Such a mean-field  approximation is probably valid for the system with localized magnetic moments. However, in case of the metallic surface the magnetic state of the impurity can be considerably suppressed by the conduction electrons of the surface. Moreover, the localization of the magnetic moments can depend on the symmetry of the $3d$ state and hybridization with the surface states. This is the case for the Co/Pt(111) system where the $yz$ and $xz$ states are itinerant and $3z^2-r^2$ is localized \cite{Mazurenko}. Thus the magnetic Hamiltonian and the conductance should contain the Heisenberg-type terms for the $3z^2-r^2$ states and Stoner-type terms for the $yz$ ($xz$) states. 
These quantum fluctuations and hybridization effects are naturally taken into account  in our Anderson model approach.

To study the interplay between transport and magnetic properties we consider the spin-matrix structure of the differential conductance,
$\frac{dI}{dV} = \sum_{\sigma \sigma'} \frac{d \mathcal{I}_{\sigma \sigma'}}{d V} $. In the spin space the tensor $d \mathcal{I}/d V$ can be expressed in the 
following form
\begin{eqnarray}
\frac{d \mathcal{I}}{d V} (\omega, \Delta, \vec{B}) = \frac{d\mathcal{I}_0}{d V} (\omega, \Delta, \vec{B}) \mathbb{1}+ \frac{d \vec {\mathcal{I}}_{sp}}{d V} (\omega, \Delta, \vec{B}) \vec \sigma,
\end{eqnarray} 
where  $\mathbb{1}$ is the unit spin matrix, $\vec \sigma$ are the Pauli matrices, $\frac{d \mathcal{I}_0}{d V} = \frac{1}{2} Tr_{\sigma} \frac{d \mathcal{I}}{d V} $ and $ \frac{d \vec{\mathcal{I}}_{sp}}{d V} = \frac{1}{2} Tr_{\sigma} \frac{d \mathcal{I}}{d V} \vec \sigma$ are spin-averaged and spin-dependent differential conductivities. Varying $\Delta$ and $\vec B$ parameters one can simulate the different tip-impurity magnetic configurations, for instance purely paramagnetic state ($\Delta$ = 0 and $\vec B = 0$), impurity paramagnetic state ($\Delta \neq 0$ and $\vec B =0$), ferromagnetic, antiferromagnetic or non-collinear states. We consider the proposed expression to be preferable in comparison with others since it preserves the spin polarization of the STM signal even when the magnetization of the adatom is small or zero. This is the case for Co/Pt(111) where $M_{Co} \sim 0.05 \mu_B$ (Fig. 2).  

In case of the collinear tip-impurity magnetic configuration, $\vec B = (0,0,B^z)$ and in the atomic limit ($U >> V_{iq}, V_{ip}$) the $\frac{d \mathcal{I}^z_{sp}}{d V}$ term can be associated with the magnetization of the individual 3d states.
The $x$-component of the differential conductance, $\frac{d \mathcal{I}^x_{sp}}{d V}$ provides the information about transitions between different spin states of the adatom \cite{Fernandez2}. In contrast to the previous works we use the electronic Hamiltonian and, therefore, varying the on-site Coulomb interaction and/or hopping integrals it is possible to study the spin dynamics of the system in the different regimes, for instance in itinerant, localized or mixed-valence states.  

We now turn to perform a quantitative analysis of the $dI/dV$ spectrum of the Co/Pt(111) system.
The components of the spin-dependent conductance calculated with different values of the tip-impurity hybridization parameters are shown in Fig. 4. One can see that in the case of weak hybridization [Fig. 4(a)] the spin-independent and longitudinal conductivities near the Fermi level are mainly due to the itinerant $yz$ and $xz$ states. The resulting spectrum demonstrates a peak at the Fermi level and we believe that the model conductance calculated with these parameters corresponds to the experimental $dI/dV$ spectrum measured by Meier $et$ $al.$ \cite{Meier}. Having calculated the conductance at temperatures ranging from 0 K to 50 K we did not find substantial changes of the spectrum. It is due to the fact that the width of the peak is much larger than the energy of the temperature excitations.

In the opposite limit of the strong hybridization (Fig. 4b) the localized $3z^2-r^2$ states give a considerable contribution to the resulting conductance and the system is in a mixed itinerant-localized state. The intensity of the conductance at the Fermi level is considerably suppressed and there is a nearly insulating behavior of the $d \mathcal{I}_0/d V$ spectrum that strongly differs from the weak hybridization case [Fig. 4(a)].   
Thus we arrive at an important conclusion that it is possible to detect the different $3d$ states of the Co impurity in the STM experiments by varying the tip-impurity distance between 4.1 \AA \, and 3.2 \AA. For this purpose the STM technique should be combined with realistic many-body computer simulation methods. 

\begin{figure}[t]
\includegraphics[width=0.4\textwidth,angle=0]{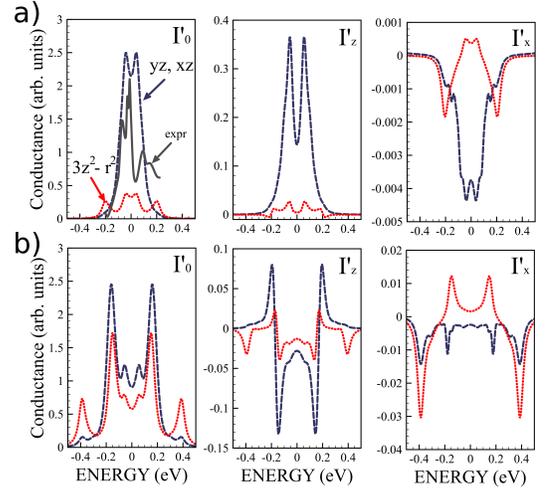}
\caption{ (Color online) The spin-independent ($\frac{d\mathcal{I}_0}{d V}$) and spin-dependent ($\frac{d \mathcal{I}^z_{sp}}{d V}$ and $\frac{d \mathcal{I}^x_{sp}}{d V}$) conductance components  calculated in different regimes with $B^z$= 8.6 T. (a) $V_{q \, 3z^2-r^2}$ = 50 meV  and $\Delta $ = 10 meV. (b) $V_{q \, 3z^2-r^2}$ = 150 meV and $ \Delta$ = 10 meV. The blue dashed and red dotted lines correspond to $yz$ ($xz$) and $3z^2-r^2$ orbitals, respectively. The grey line is the experimental $dI_{STM}/dV$ spectrum taken from Ref.\cite{Meier}.}
\label{structures}
\end{figure} 

It is also interesting to analyze the transverse component of the conductance that is originated from the spin-flip excitation, for instance, $\langle n | c^{+}_{p \uparrow} d_{i \uparrow} d^{+}_{i \downarrow} b_{q \downarrow}  | n \rangle$. The fact that using a Schrieffer-Wolff transformation the electronic Hamiltonian Eq.(1) can be reduced to an effective tunneling spin Hamiltonian \cite{Appelbaum, Anderson, fernandez} containing the spin-flip processes confirms that the original Hamiltonian also provides such excitations. The calculated $yz$ ($xz$) and $3z^2-r^2$ transverse contributions demonstrate remarkably different behavior. While the $\frac{d \mathcal{I}^x_{sp}}{d V}$ spectrum of the itinerant states is negative, the localized curve changes the sign near the Fermi level. The conductivities calculated in weak and in strong hybridization regimes are mainly associated with the $yz$ and $3z^2-r^2$ states, respectively. The intensity of the spin-flip excitations varies strongly depending on the tip-impurity distance. At  $V_{q \, 3z^2-r^2}$ = 150 meV such excitations strongly contribute to the resulting conductance since the system becomes more localized and insulating. 
 
In conclusion, we numerically demonstrate that the differential conductance of the Co/Pt(111) system measured in the STM experiments is orbital-selective. Varying tip polarization and tip-impurity distance one can get the access to the localized and itinerant states of the Co impurity. The current SP-STS experiments operate with itinerant $yz$ and $xz$ states demonstrating a weak response to the applied magnetic field. 
The obtained results are useful for future scanning tunneling experiments aiming to manipulate individual $3d$ states of a transition metal adatom deposited on a surface.  

{\it Acknowledgements.}
We thank M.I. Katsnelson, V.I. Anisimov, T. O. Wehling, A. I. Poteryaev  and A.A. Katanin for helpful discussions.
The hospitality of the Institute of Theoretical Physics of Hamburg University is gratefully acknowledged.
This work is supported by DFG Grant  No. SFB 668-A3 (Germany), RFBR 10-02-00546-a, the grant program of President of Russian Federation
 MK-406.2011.2, Intel Scholarship Grant, the scientific program ``Development of scientific potential of universities'' N 2.1.1/779 and by the scientific program of the Russian Federal Agency of Science and Innovation N 02.740.11.0217.

\end{document}